# What Causes My Test Alarm?
# Automatic Cause Analysis for Test Alarms in System and Integration Testing

He Jiang[1,2,4] Xiaochen Li[1] Zijiang Yang[3] Jifeng Xuan[4]

[1]School of Software, Dalian University of Technology, Dalian, China
[2]Key Laboratory for Ubiquitous Network and Service Software of Liaoning Province, Dalian, China
[3]Western Michigan University, Kalamazoo, MI, USA
[4]State Key Lab of Software Engineering, Wuhan University, Wuhan, China
jianghe@dlut.edu.cn li1989@mail.dlut.edu.cn zijiang.yang@wmich.edu jxuan@whu.edu.cn

**Abstract—Driven by new software development processes and testing in clouds, system and integration testing nowadays tends to produce enormous number of alarms. Such test alarms lay an almost unbearable burden on software testing engineers who have to manually analyze the causes of these alarms. The causes are critical because they decide which stakeholders are responsible to fix the bugs detected during the testing. In this paper, we present a novel approach that aims to relieve the burden by automating the procedure. Our approach, called Cause Analysis Model, exploits information retrieval techniques to efficiently infer test alarm causes based on test logs. We have developed a prototype and evaluated our tool on two industrial datasets with more than 14,000 test alarms. Experiments on the two datasets show that our tool achieves an accuracy of 58.3% and 65.8%, respectively, which outperforms the baseline algorithms by up to 13.3%. Our algorithm is also extremely efficient, spending about 0.1s per cause analysis. Due to the attractive experimental results, our industrial partner, a leading information and communication technology company in the world, has deployed the tool and it achieves an average accuracy of 72% after two months of running, nearly three times more accurate than a previous strategy based on regular expressions.**

## I. INTRODUCTION

System and Integration Testing (SIT) is necessary immediately after the integration of various software components. With increasing number of companies advocating to conduct continuous integration [32] by following modern software development practices such as DevOps [31], the frequency of SIT has significantly increased. Fortunately, emerging techniques such as testing in the cloud have dramatically improved the efficiency of such testing. For example, a cloud-based system is able to run 1,000 test scripts in less than 25 minutes. In the past the same amount of testing required 77 hours [33]. Since running test scripts has an average failure rate of approximately 5% [14], the frequent automated SIT produces tremendous number of test alarms that have to be analyzed by testers.

There are various causes that may lead to test alarms, such as product code defect, test script defect and device anomaly. Each type of cause has its unique way to handle, including submitting bug reports to developers, correcting the test scripts and submitting exception messages to instrument suppliers. Therefore, the analysis of test alarms is critical as it determines who is responsible to fix the potential bugs.

In order to figure out the causes, testers have to carefully read test logs [20], each of which may consist of hundreds of test steps and thousands of lines of text [4]. Considering the fact that thousands of test alarms may be produced per day for a production line with several similar products, as we have observed during collaboration with our industrial partner Huawei-Tech Inc., a leading information and communication technology company in the world, test alarm cause analysis lays an almost unbearable burden on testers and has become a bottleneck in SIT. Realizing the urgent need to alleviate the burden of cause analysis, our collaborators manually build regular expressions over the test logs to analyze test alarm causes. The accuracy of their approach is about 20%-30% on different projects.

In this paper, we present a novel approach named Cause Analysis Model (CAM) that infers test alarm causes by analyzing test logs. The test logs, generated by test scripts, record important runtime information during testing. The basic idea of CAM is to detect the test logs of historical test alarms that may share the same causes with the new test logs. CAM first utilizes Natural Language Processing (NLP) techniques to partition test logs into terms. Next CAM selects partial historical test logs for further processing with function point filtering. Thirdly, CAM constructs attribute vectors based on test log terms. The cause of a new alarm is predicted according to the ranked similarity between a new test log and each historical one. Finally, CAM reports the causes along with the difference between the new and historical test logs. CAM is efficient as it is an information retrieval based algorithm without the overhead of training.

In the experiments, we collect more than 14,000 test logs, forming two datasets, from two industrial projects at Huawei-Tech Inc. CAM achieves accuracy rates of 58.3% and 65.8%, respectively, outperforming baseline algorithms by up to 13.3%. For more than one-third of the testing days, the accuracy of CAM is over 80%. In addition, CAM is very efficient, taking on average about 0.1s per test alarm analysis with 4GB memory. After deploying CAM at Huawei-Tech Inc., it achieves an average accuracy of 72% after two months of running, which is nearly three times more accurate than their previous strategy based on regular expressions.

In summary, this study makes the following contributions:

(1) We propose a new approach to address the challenge of automatically analyzing the test alarm causes in SIT.

(2) We construct two industrial datasets with more than 14,000 test logs. The failure causes of these test alarms are manually labeled and verified by testers.

(3) We conduct a series of experiments to investigate the performance of our approach. Experimental results show that CAM is both effective and efficient.

(4) We deploy and evaluate CAM at Huawei-Tech Inc. in a real development scenario.

This paper is structured as follows. In Section 2, we introduce the background of this study. We describe the overall framework of CAM in Section 3. The experimental setup and research questions are introduced in Section 4. We experiment to answer the research questions in Section 5. Section 6 and 7 show the threats to validity and related work, respectively. Finally, Section 8 concludes this paper.

## II. BACKGROUND

In this section, we present relevant background regarding system and integration testing and the cause analysis problem.

### A. System and Integration Testing (SIT)

*SIT* is performed immediately after various components are integrated to form an entire system. The system under test is more complex than those individual components considered in the unit testing. Therefore, SIT uses a new set of test drivers for revalidation with black-box testing strategies [36].

The *function points* are a set of functional requirements for a software project [46]. In SIT, testers play the role of users to work through a variety of scenarios for covering the required function points [45]. The function points of test scripts are predefined when testers develop test scripts [45]. For example, if a test script is designed to verify the function of "configure network proxy", testers may add "NETCONF_PROXY _FUNC" as the function point of the test script.

*Test logs* record the runtime information in software testing. In SIT, testers develop test scripts (also called test codes [34]) to check for system functions, performance, etc. Each test script contains a sequence of test steps with numerous logging statements. Test logs are generated by these logging statements when running test scripts.

A *test alarm* is an alarm to warn the failure of a test script. Each test alarm is associated with a failure cause. Testers are responsible to analyze the causes of test alarms.

### B. Cause Analysis Process

Cause analysis for test alarms is critical due to its effect on both testers and developers [4]. The overall analysis procedure is depicted in Fig. 1. In a software company, SIT is conducted over the code changes in each branch to reduce software bugs [4]. Before developers merge code changes into a trunk branch, testers select test scripts of some given function points to verify the correctness of these code changes (Fig. 1(1)). During the testing, test scripts automatically log important runtime information to form test logs. Code changes are merged into a trunk branch only if they pass all the test scripts. If a test script fails, testers are required to analyze the cause to the failure (Fig. 1(2)).

Testers analyze failure causes by examining test logs (Fig. 1(2)). After detecting failure causes, testers submit the test logs with the corresponding causes to the software repository for unified management (Fig. 1(3)). After that, different stakeholders, e.g., testers, developers, instrument suppliers, etc., have to resolve the failures depending on the types of the causes (Fig. 1(4)). If a cause indicates product code defect, testers need to submit a bug report to developers and request them to fix the bug [51]. If it is a defect in test scripts, testers need to correct test scripts. For other causes, testers may either adjust the configuration files, or request instrument suppliers to diagnose the infrastructures, etc. The above process may repeat several times before code changes are merged into a trunk branch (Fig. 1(5)).

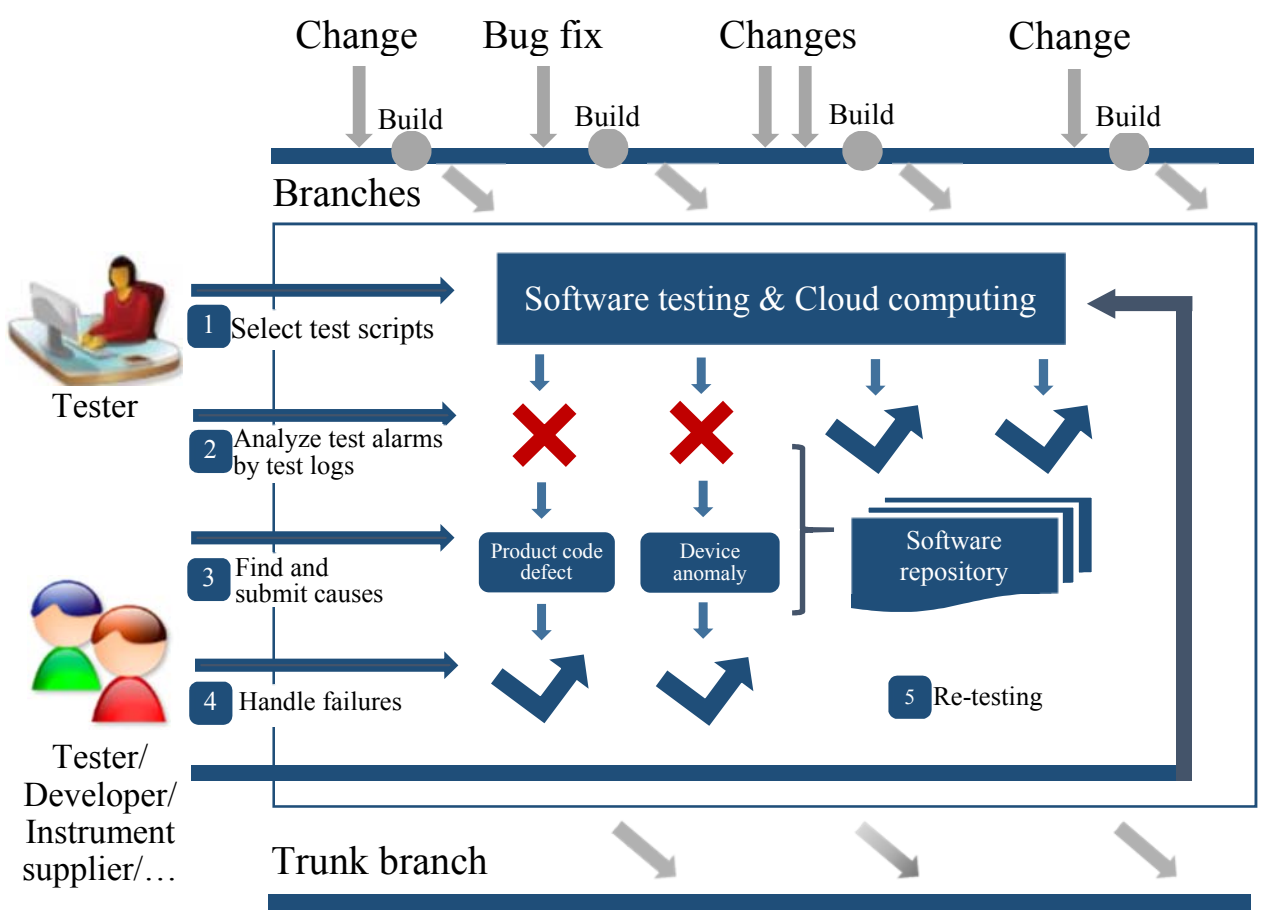

**Figure 1. Cause analysis process.**

### C. Cause Analysis Problem

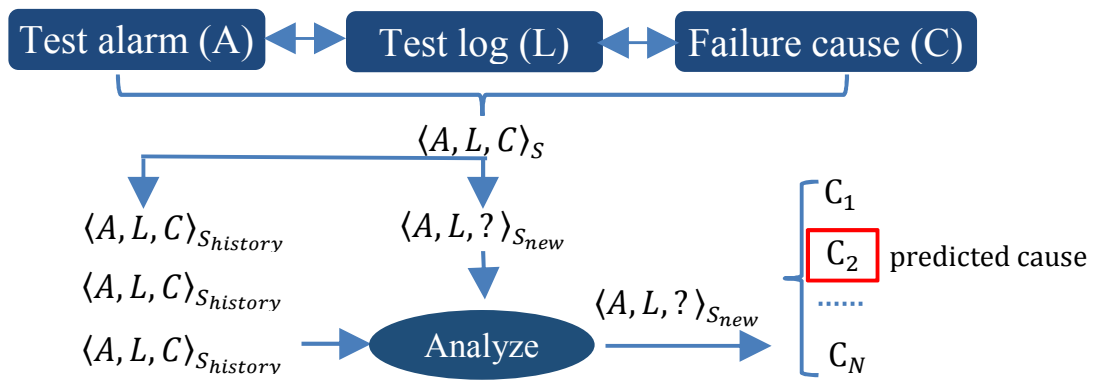

**Figure 2. Cause analysis problem.**

As shown in Fig. 2, each test alarm (A) is associated with a test log (L) and its failure cause (C), which forms a triple $\langle A,L,C\rangle$. When a set of test scripts completes running, several $\langle A,L,?\rangle_S$ arise for analysis. Testers analyze the causes of test alarms with their test logs, and then continuously submit the $\langle A,L,C\rangle_S$ to the software repository along with the testing days.

We represent $\langle A,L,C\rangle_S$ for analysis as $\langle A,L,?\rangle_{S_{new}}$, and those in the software repository as $\langle A,L,C\rangle_{S_{history}}$. Following this representation, the cause analysis problem is to predict C in $\langle A,L,?\rangle_{S_{new}}$ with the assistance of $\langle A,L,C\rangle_{S_{history}}$, which can be viewed as a multiclass classification problem due to the various failure causes (C) for test alarms. The multiclass classification problem aims to classify instances into one out of more than two classes. In this study, the new test logs of test alarms are instances for classifying, and their causes are the multiple classes.

Despite previous studies [4] [34] attempt to classify test alarms into product code defect and non-product code defect, these techniques are not suitable for this problem, since they either require expensive costs to collect complex information in large integrated system [34] or need additional efforts to decide how to deal with each non-product code defect [4].

### D. Test Logs and Failure Causes

We exhibit some examples of test logs and failure causes from the industrial projects at Huawei-Tech Inc. to better understand the cause analysis problem. The projects are launched to test the codes of two communication systems.

*1) Test Logs*

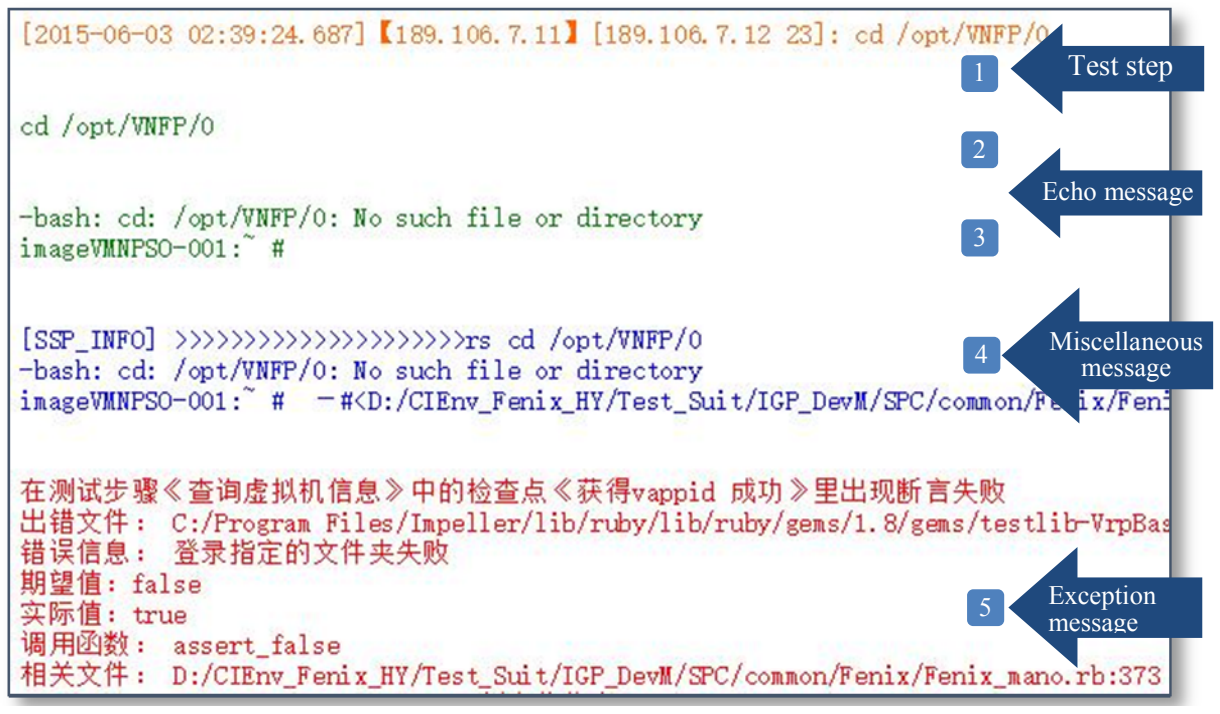

**Figure 3. A snippet of test logs.**

Logging is a conventional programming practice to record important runtime information [1]. In some open-source software, there is one line of logging code in less than 50 lines of code on average [2]. When testers develop test scripts, they also insert a mass of logging statements [3]. During the runs of test scripts, these logging statements record some critical information to test logs.

Fig. 3 exhibits a snippet of test logs. Test logs in these projects are bilingual documents with English and Chinese terms. In practice, non-English speaking testers prefer adding some native terms, e.g., Chinese terms, to better understand test logs. Apart from the languages, test logs in these projects contain all the information that a test log needs [3].

The contents of the industrial test logs can be summarized in four types, including test steps, echo messages, exception messages, and miscellaneous messages.

A *test step* (segment 1) is a command or code snippet to display or verify some specific steps of the software under test. A test script contains a sequence of test steps, simulating the operations of a user to cover the required function points.

An *echo message* (segments 2 and 3) is a feedback of the test step, which may contain output actions, state of object, environment variables, etc.

*Exception messages* (segment 5) record the critical information when a test script fails, which often contain the functions or files being called during the test alarm.

All the segments except the test steps, echo messages, and exception messages, are classified as *miscellaneous message* (segment 4), which may include prompt messages and messages from related infrastructures.

In conclusion, test logs record information about testing activities, including the state of test scripts, the software under test, and related infrastructures, etc. However, it is a non-trivial work to fully distinguish all the information, since the distribution of the information varies over distinct projects. Testers peruse the entire test logs to analyze testing activities.

*2) Failure Causes*

Table 1 exhibits the explanations of the test alarm causes in the two projects. We also present the solutions to these test alarms, namely how testers deal with each test alarm.

**Table 1. Causes for test alarms and solutions**

| ID | Type of cause | Explanation | Solution |
|---|---|---|---|
| C1 | *Obsolete test* | Test scripts or product codes are obsolete when continuous integration, e.g., testers conduct testing with out-of-date test scripts. | Testers update test scripts or product codes. |
| C2 | *Product code defect* | Defects in product code, e.g., the product code does not meet the requirement of a function point. | Testers submit bug reports to developers |
| C3 | *Configuration error* | Configuration files are incorrectly edited, e.g., testers set conflict parameters in configuration files. | Testers correct configuration files |
| C4 | *Test script defect* | Faults in assertion expression, arguments, statement of test scripts, e.g., quotation marks mismatch in test script. | Testers debug test scripts |
| C5 | *Device anomaly* | Defects exist in the devices for running the test bed, e.g., the interface board of running the communication system breaks down. | Testers submit bug reports to instrument suppliers |
| C6 | *Environment issue* | Environment issues include the problems of the network, CPU, memory, etc., e.g., the space of hard disk is not enough for executing test scripts. | Testers diagnose the environment |
| C7 | *Third-party Software problem* | Defects or incompatible issues exist in the third- party software, e.g., there are problems for the automatic testing system. | Testers ask site reliability engineers to diagnose the third-party software |

There are seven types of causes in the projects. We find that handling test alarms in SIT is a complex process. On the one hand, different causes lead to distinct solutions. Debugging or locating bugs in test scripts (C4) is not enough for testers to handle test alarms. Testers may conduct obsolete test (C1), wrongly configure some files (C3), or face several environment issues (C6), etc. On the other hand, testers also need to cooperate with distinct stakeholders to handle test alarms. Testers send all the product code defects (C2) to developers. Some device anomalies (C5) also require the instrument suppliers to deal with. Site reliability engineers are responsible for fixing third-party software problems (C7).

Hence, automatically deciding the type of causes can help testers focus on some specific resources. For example, if it is already known that a test alarm is caused by the test script defect, testers can further run some bug location and fixing tools for deeper analysis.

In addition, many types of causes in Table 1 also exist in open-source software. After investigating the causes for false test alarms (all test alarms caused by non-product code defects) of Apache software [35], we find that causes C1, C3, C4, and C6 are also detected in [35].

## III. CAUSE ANALYSIS MODEL

In this section, we present our Cause Analysis Model (CAM) in detail. The basic idea of CAM is to search the test logs of historical test alarms that may have the same failure cause with the new test log. As shown in Fig. 4, CAM first pre-processes test logs with bilingual NLP techniques. Then, historical test logs are selected according to the function points. Third, CAM predicts the cause of a new test alarm based on similarity between new and historical test logs. Finally, both the cause and the difference between new and historical test logs are presented to facilitate the examination of prediction results.

CAM is efficient as it is an information retrieval based algorithm without the overhead of training models. Besides,

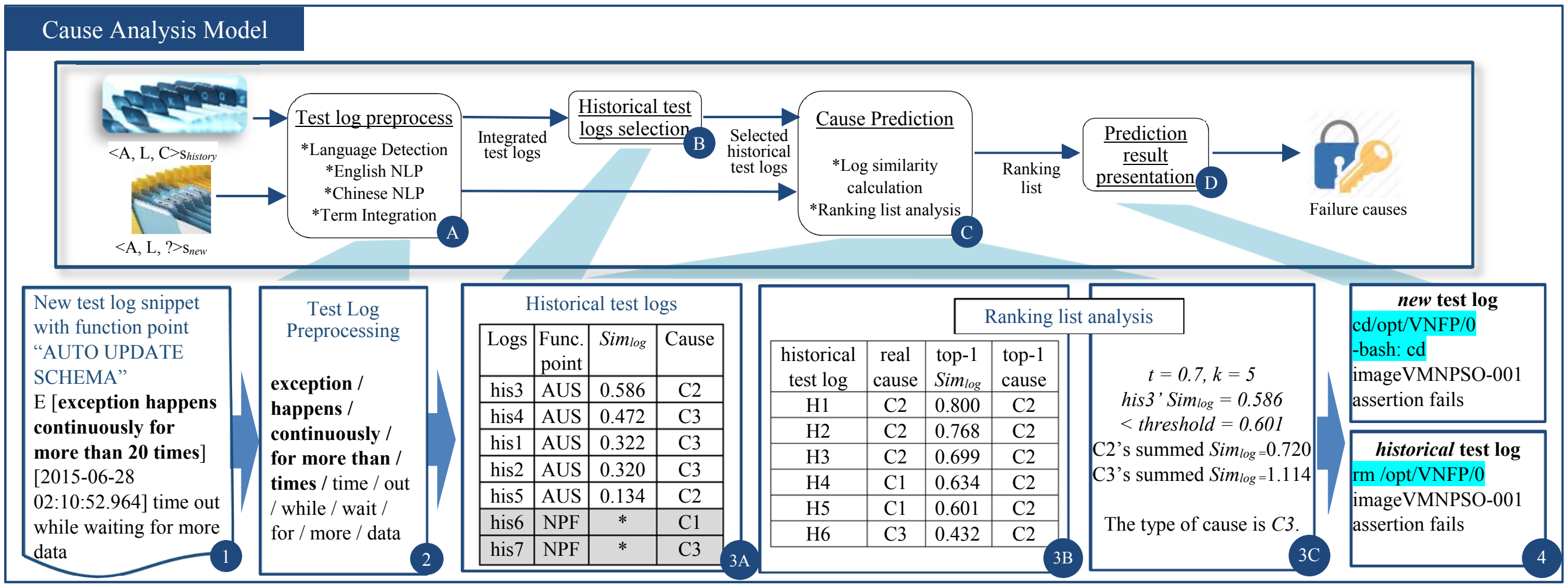

**Figure 4. Framework of CAM for test alarm analysis with a running example**

testers could better understand and verify the prediction results after examining the information presented by CAM.

We exhibit a running example to predict the cause of the test log snippet in Fig. 4(1). The test log is generated by a test script for verifying the function point "AUTO_UPDATE _SCHEMA" (AUS for short). The test log shows that "time out while waiting for more data". In addition, testers use some Chinese messages to warn that "exception happens continuously for more than 20 times". We translate and present the Chinese part in bold.

## A. Test Log Preprocessing

In this study, test logs are bilingual documents, which makes test log preprocessing more complex than that in a single language. CAM preprocesses these test logs with a series of bilingual NLP techniques. For test logs written in merely English, only English NLP is needed.

**Language Detection**. We first distinguish the texts in a test log by the language type. Since languages are located at independent areas in the UTF-8 encoding table, we apply a regular expression "[\\u4e00-\\u9fa5]+" to detect the areas. Terms matching the regular expression are classified as Chinese terms. The remaining ones are English terms. Therefore, a test log is separated into two independent parts, namely the English part and the Chinese part.

**English NLP**. We apply three widely used English NLP steps to preprocess the English part, including tokenization, stop words removal, and stemming. First, the English part is tokenized with a regular expression "[\w-]+(\.[\w-]+)*" [38]. String meets the regular expression is tokenized as a term. Second, we consider the single letter terms, punctuation marks, and numbers as stop words and remove them, e.g., 'E' and "2015-06-28" in Fig. 4(1). Third, porter-stemming algorithm [39] is employed to stem each term.

**Chinese NLP**. Word segmentation is a major difference between Chinese and English NLP steps, since Chinese documents are written without any spaces between terms [7]. We utilize IKAnalyzer, an open source NLP tool [41], for Chinese word segmentation. In Fig. 4(2), IKAnalyzer detects several terms in the Chinese part, including, "exception / happens / continuously / for more than / times".

**Term Integration**. After transforming the test log into terms, we merge English and Chinese parts together according to their original order for unified operations (in Fig. 4(2)).

## B. Historical Test Log Selection

Historical test log selection aims to select a subset of historical test logs for efficient cause prediction, as it is time consuming and noise overwhelming to search through all the historical test logs. For example, there are seven historical test logs in Fig. 4(3A). After selection, only test logs "his1" to "his5" are used for cause analysis.

CAM selects historical test logs by examining the function points of test scripts, since we find that test scripts with the same function point usually target the same functionalities to check. They are more likely to fail with the same cause as we investigate in section V(D). For a project under test, test specifications [47] can be an effective material to extract the function points of test scripts.

In this study, we extract the function point of each test script directly from the automatic testing system of our industry partner. We associate the function points with test logs by matching the test script ID. For example, in Fig. 4(3A), the function points of test logs "his1" to "his5" are "AUS", while that of "his6" and "his7" are "NPF" ("NETCONF_ PROXY_FUNC"). When analyzing the test log of a new test alarm, CAM attempts to select historical test logs for the same function point as the new one. If no such historical test log is available, CAM utilizes all the historical test logs to conduct the prediction.

## C. Cause Prediction

After historical test log selection, CAM predicts the cause of a new test alarm by first ranking the selected historical test logs according to their similarities with the new test log, and then analyzing the ranking list to achieve the possible cause. The basic hypothesis is that the possibility of two test logs implying the same cause increases along with the growth of the similarity between two test logs.

### 1) Log similarity calculation

CAM calculates the similarity between test logs by cosine similarity measurement [7]. The inputs of this measurement are the attribute vectors of test logs. We construct attribute vectors with the 2-shingling strategy, in which each 2-shingling is an attribute. More specifically, if we view a document as a sequence of terms, a contiguous subsequence in the document is referred to as a shingle. The 2-shinglings are defined as the set of all unique shingles of size 2 in this document [6]. The 2-shingling strategy has been successfully applied in Chinese grammar detection [8], information retrieval [7], etc. For example, "exception happens" and "happens continuously" are two 2-shinglings of the snippet in Fig. 4(2). CAM calculates the weight of attributes with TF-IDF (Term Frequency-Inverse Document Frequency) [7]. For an attribute $A$ in the test log $T$, its TF-IDF value is defined as:

$$\text{TF-IDF}_{A,T} = f_{A,T} * \log\frac{N}{n_A} \quad (1)$$

, where $f_{A,T}$ denotes the number of times that $A$ occurs in $T$, $n_A$ denotes the number of test logs containing $A$, and $N$ denotes the number of test logs in a project.

With the attribute vectors of two test logs, the cosine similarity [7] is measured as:

$$Sim_{log}(\vec{V}1, \vec{V}2) = \frac{\vec{V}1 \cdot \vec{V}2}{|\vec{V}1||\vec{V}2|} \quad (2)$$

, where $\vec{V}1 \cdot \vec{V}2$ is the inner product of two vectors and $|\vec{V}1||\vec{V}2|$ is the product of 2-norm for these vectors.

CAM ranks the selected historical test logs in descending order by $Sim_{log}$. In Fig. 4(3A) the test log "his3" has the largest $Sim_{log}$ = 0.586 and its cause is "C2".

*2) Ranking list analysis*

It is reasonable to predict the test alarm cause with the top-1 cause[1] in the ranking list. However, we find that the top-1 $Sim_{log}$ may be very low in some ranking lists, which can be interpreted as that the top-1 test log has a small possibility to share the same cause with the new one. Considering such a situation, CAM analyzes the test alarms as follows. For a new test log, if its top-1 $Sim_{log}$ is larger than a cause-specific *threshold*, CAM adds the new test log to a high-similarity set. Otherwise, CAM adds it to a low-similarity set. For a test log in the high-similarity set, CAM predicts the new log's cause as the top-1 cause in the ranking list. For a test log in the low-similarity set, a KNN [50] strategy is conducted.

**Algorithm 1: Cause-specific threshold calculation**

**Input**: the set D of all historical test logs with the top-1 cause $i$;
the target value $t$
**Output**: threshold $\theta_i$

```
let threshold θi = 1.0
for x= 0; x<=1; x=x+0.001 do
    achieve the test logs with the top-1 Sim_log > x from D
    if the portion of the above test logs with the real cause i > t.
        threshold θi = x; break;
    else
        continue;
end
return threshold θi
```

**Cause-specific threshold**. We employ a cause-specific threshold $\theta_i$ for cause $i$ to differentiate the high-similarity set and the low-similarity set. Since we could achieve a ranking list for a historical test log by calculating the $Sim_{log}$ for every test log prior to this historical test log, each historical test log is associated with a top-1 cause and top-1 $Sim_{log}$. Given a new test log with top-1 cause $i$, the value of $\theta_i$ is defined as the minimum value between 0 and 1, such that over $t$*100% of the causes of all the historical test logs with the top-1 cause $i$ and the top-1 $Sim_{log} > \theta_i$ are correctly predicted as cause $i$, where $t$ is a parameter named target value (the impact of $t$ is discussed in Section V). In algorithm 1, we elaborate the details on how to determine the threshold $\theta_i$ for cause $i$.

Taking Fig. 4(3B) as an example, there are six historical test logs with the top-1 cause C2 in the set D. We increase $x$ from 0 to 1. When $x$=0.500, only H1-H5 are considered. The top-1 causes of H1-H3 equal the real cause, therefore the portion is 0.6. When $x$=0.601, only H1-H4 are considered, among which the portion reaches 0.75, larger than the target value 0.7. At last, the threshold of cause C2 is 0.601.

**KNN strategy**. For a test log in the low-similarity set, we sum up the $Sim_{log}$ values of the top-k test logs by their causes. Then, the cause of the new test log is assigned to the cause with the largest summed $Sim_{log}$.

As the $Sim_{log}$ of "his3" is smaller than C2's threshold 0.601, CAM obtains top-$k$ test logs in the ranking list (Fig. 4(3C)). After being summed, cause C3 achieves the largest summed $Sim_{log}$ =1.114, which are summed by the $Sim_{log}$ of "his4", "his1", and "his2". Hence, CAM predicts the cause of the new test log as C3.

## *D. Prediction result presentation*

CAM helps testers understand and verify the prediction results by presenting the causes as well as the differences between test logs. For test logs in the high-similarity set, CAM shows the differences between the new and the top-1 test log in the ranking list. For test logs in the low-similarity set, CAM shows the differences between the new test log and the first historical test log with the predicted cause, e.g., "his4" in Fig. 4(3A). Since historical test logs have been analyzed by testers, testers may easily know whether two test logs implying the same failure causes after perusing the differences.

To show the differences, CAM first removes all the numbers in test logs, since such information usually indicates time, IP address, and numeric counter, etc., which may be different in all the test logs. Then, CAM compares the differences between test logs with "JavaDiffUtils", an open source tool to compare differences between texts [40]. The tool shows all the "Change", "Delete", and "Insert" operations between texts. At last, we highlight the different lines of the two test logs. For the "Change" operations, the lines in both two test logs are highlighted. For the "Delete" or "Insert" operations, only the lines with more information are highlighted.

For example, in Fig. 4(4), only the first two lines are different. Instead of comprehending the entire contents of the test log, testers can focus on the first two lines to verify the prediction result with the assistance of the historical test log.

# IV. EXPERIMENTAL SETUP

In this section, we detail the experiment related issues for evaluating CAM. The datasets and evaluation metrics are first

[1] The top-1 cause refers to the top-1 test log's cause, while the top-1 $Sim_{log}$ refers to the top-1 test log's $Sim_{log}$ in a ranking list.

presented, followed by a discussion of the baseline algorithms and the Research Questions (RQs).

## A. Datasets

**Table 2. Statistic of test logs and causes in the datasets**

| # | Info \ Dataset | DS1 | | DS2 | |
|---|---|---|---|---|---|
| 1 | **# Test logs** | 7663 | | 6977 | |
| 2 | **Size** | 4.72GB | | 6.06GB | |
| 3 | **Time Frame** | June 1st – July 30th, 2015 | | Oct. 26th – Nov. 16th, 2015 | |
| 4 | **# Testing day** | 40 day | | 22 day | |
| 5 | **# Test logs per day** | 192 | | 317 | |
| 6 | **# Avg. lines** | 942 lines | | 1375 lines | |
| 7 | **# Avg. test steps** | 247 test steps | | 344 test steps | |
| 8 | **# Obsolete test (C1)** | 1185 | 15.46% | * | * |
| 9 | **# Product code defect (C2)** | 4459 | 58.19% | 1963 | 28.14% |
| 10 | **# Configuration error (C3)** | 761 | 9.93% | 345 | 4.94% |
| 11 | **# Test script defect (C4)** | 892 | 11.64% | 3259 | 46.71% |
| 12 | **# Device anomaly (C5)** | 335 | 4.37% | 298 | 4.27% |
| 13 | **# Environment issue (C6)** | 19 | 0.28% | 168 | 2.41% |
| 14 | **# Software problem (C7)** | 12 | 0.17% | 944 | 13.53% |
| 15 | **# Avg. type of causes per day** | 3.85 per day | | 3.86 per day | |

We collect test logs from two industrial testing projects at Huawei-Tech Inc. to build two datasets, denoted as DS1 and DS2. In the datasets, each test log, corresponding to a test alarm, is associated with a failure cause manually labeled by the testers. Table 2 exhibits the statistical information of the datasets. Rows 1 to 7 show that there are more than 14,000 test logs, including 7663 for DS1 and 6977 for DS2. These test logs are generated during 40 and 22 valid testing days, due to vacations and other testing activities. On average, testers are requested to analyze 192 and 317 test alarms per day in the two projects. As shown in rows 6 and 7, each test log contains more than 900 lines of texts, including about 247 and 344 test steps. The total size of the datasets exceeds 10GB. Hence, test logs are a relatively large software artifact, which may consume considerable time for analysis. Rows 8 to 14 present the number of test logs with respect to each type of cause. There are 7 and 6 types of causes detected in DS1 and DS2, respectively. *Obsolete test* (C1) never occurs in DS2 during the time frame.

Based on the statistic information, we have the following observations.

(1) Besides *Product code defect* (C2) and *Test script defect* (C4), other causes still cover 30% of test alarms. As shown in row 15, nearly four types of causes occur per day on average. These factors make testers have to decide the exact type of cause for each test alarm before fixing these alarms.

(2) Another finding is that, no test alarm is caused by more than one type of cause in the datasets. One possible reason is that the causes defined in this study are for classifying the candidate part with defects, e.g., devices, environment, etc. A single buggy part can directly lead the test script to fail. In addition, after studying the daily work of testers in the projects, we find that if testers confirm the part causing the test alarm, they seldom diagnose the remaining parts, unless the test script fails again. Thus, automatically identifying the most possible cause for test alarms is beneficial for testers.

## B. Evaluation Method

We utilize the incremental framework [42] to run algorithms in all the experiments, which can better simulate the daily work of a tester. More specifically, we partition the datasets by the testing day. When analyzing the test logs in $day_T$ ($T \geqslant 1$), we regard test logs in $day_0$ to $day_{T-1}$ as historical test logs. Since the framework conducts prediction from the second day, there are 39 and 21 testing days with 7356 and 6557 test logs in the datasets to be predicted respectively.

Under the incremental framework, we evaluate the overall performance of algorithms with Accuracy and AUC (Area Under roc Curve).

$$Accuracy = \frac{Num_{correct}}{Num_{analyzed}} \quad (4)$$

Accuracy can be interpreted as the portion of correctly predicted test logs among all the predicted ones.

$$AUC_i = \int_{\infty}^{-\infty} TPR_i(T)FPR_i'(T)dT,$$

$$where\ TPR_i = \frac{\#positives_i\ correctly\ classified}{\#total\ positives_i},$$

$$FPR_i = \frac{\#negatives_i\ incorrectly\ classified}{\#total\ negatives_i} \quad (5)$$

AUC is the area of the two-dimensional graph in which $TPR_i$ is plotted on the Y axis and $FPR_i$ is plotted on the X axis over distinct threshold $T$ of possibility values [28]. A possibility value of CAM's prediction is the $Sim_{log}$ of the first historical test log with the predicted cause in a ranking list. We calculate AUC in a one against all class strategy. When calculating AUC of cause $i$, all the test alarms predicted as causes $i$ are considered as $positives_i$, while the other types of causes are considered as $negatives_i$. AUC can avoid inflated performance evaluation on imbalanced data. For example, a classifier that always predicts "Product code defect (C2)" achieves 58.19% accuracy on DS1, but results in an AUC of 50%, which is the same AUC as a random guess classifier.

In addition, we also evaluate the resource consumption of these algorithms. Resource consumption is critical in industry projects. On the one hand, computation resources, e.g. CPU, memory, are limited in real scenarios. On the other hand, testers may conduct software testing several rounds per day. Low resource consumption makes algorithms timely update models with the information in the latest round. Thus, we evaluate the time and minimal memory for completing the prediction of each algorithm.

## C. Baseline Algorithms

To the best of our knowledge, no studies directly utilize test logs to predict the causes of test alarms in SIT. We implement three baseline algorithms to study the characteristics of CAM.

**Lazy Associative Classifier** (LAC). A similar work by Herzig et al. detects false test alarms with association rules mined from test steps [4]. Since the mining algorithm in [4] is not suitable for multiclass classification, we employ lazy associative classifier to predict the causes, which uses association rules to execute multiclass prediction [43]. Following the strategies in [4], we extract test steps from test logs. In our datasets, test steps can be identified since they are marked with timestamps. We build attribute vectors for LAC based on the test steps. Each entry in the vector represents whether a test step exists in a test log. The parameters of association rules, namely minimum *confidence* and *support* values, are set to 0.8 and 0.03 respectively [4]. We implement LAC with an open source tool shared by Federal University of Minas Gerais [44].

**Best First Tree** (BFT). Hao et al. [34] classify test alarms into *product code defect* and *obsolete test script* at the unit testing stage with BFT classifier. Since the attributes related to test complexity and program execution measurements are expensive to collect in large software systems [4], we examine whether the BFT classifier is suitable for cause analysis. BFT is implemented with the widely used machine learning tool WEKA [37]. We alternatively use the TF-IDF values of terms in a test log as attributes for BFT's input.

**Topic Model** (TM). As a popular way to analyze a large scale of documents, TM can be used to predict the test alarms causes. TM first extracts several topics from test logs by mining co-occurrence terms. Next, each test log is expressed by a series of topics with different probabilities. Thirdly, we construct attribute vectors with these topics and utilize the cosine similarity measurement to rank historical test logs. At last, the top-1 cause in the ranking list is used for the prediction. We implement one type of TM, namely Latent Dirichlet Allocation, with an open source tool Mallet [30]. We set the parameter of topic number to 200, alpha to 0.01, and beta to 0.01 according to the suggestion by Mallet.

In the experiments, we set the test logs in $day_0$ as the initial training set, and incrementally train models after each testing day, such that these algorithms can fully learn all the information from history.

### D. Research Questions

**RQ1**: Are the test logs with the same causes more similar than those with different causes?
**RQ2**: How do the parameters influence CAM's performance?
**RQ3**: How does CAM perform against baseline algorithms?
**RQ4**: How does historical test log selection influence the performance of CAM?
**RQ5**: How does CAM perform in a real development scenario?

## V. EXPERIMENTAL RESULTS

All the algorithms are implemented in Java JDK1.8.0_31, and run on a PC with Intel Core(TM) i7-4790 CPU 3.6GHz and 24G memory.

### A. Answer to RQ1

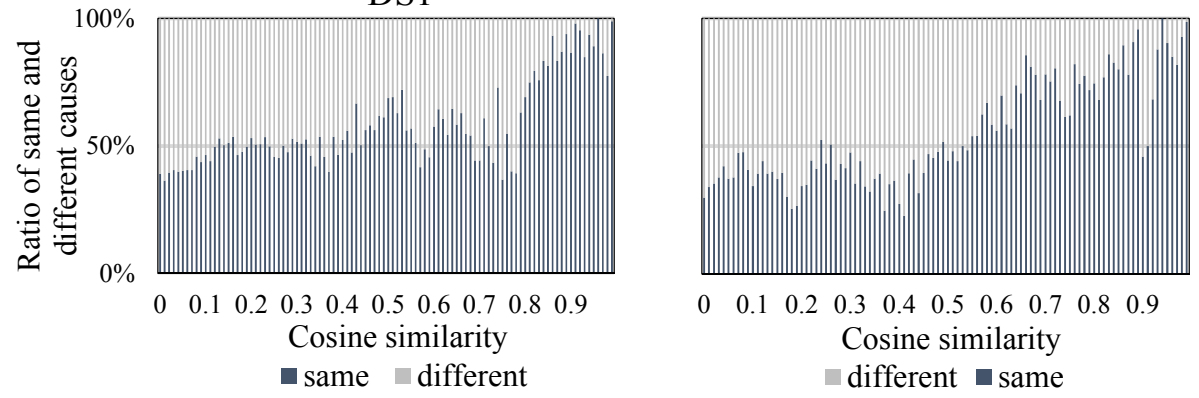

**Figure 5. Causes vs. cosine similarity of test logs.**

CAM predicts test alarm causes by the similarity between test logs. In this RQ, we verify CAM's hypothesis, namely, test logs with the same causes are more similar than those with different causes.

We calculate the pairwise $Sim_{log}$ of test logs on the two datasets and collect all the test log pairs with $Sim_{log}$ between $x$ to $x$+0.01, where $x$ ranges from 0 to 0.99 with a step size of 0.01. In Fig. 5, the dark blue part of a bar in the bar chart presents the ratio of test log pairs with the same causes in distinct similarity range. As shown in Fig. 5, the ratio of test logs with the same causes gradually increases along with the increment of the $Sim_{log}$. More than 50% of test log pairs are with the same causes when $Sim_{log} > 0.79$ on DS1 and $Sim_{log} > 0.55$ on DS2. Test logs with the same causes tend to have a higher $Sim_{log}$ than those with different causes.

**Answer to RQ1**. The possibility of two test logs implying the same causes increases along with the growth of the similarity between two test logs. Test logs with the same causes are more similar than those with different causes.

### B. Answer to RQ2

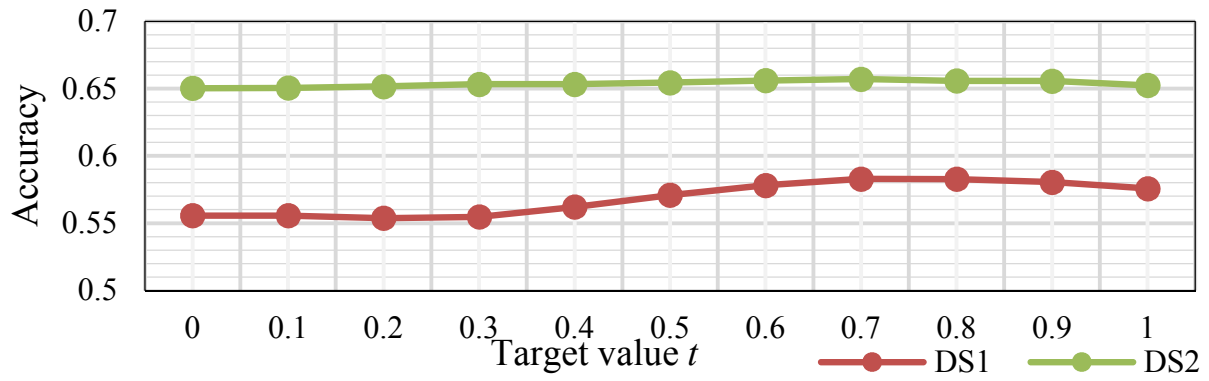

**Figure 6. Accuracy with varied target value *t*.**

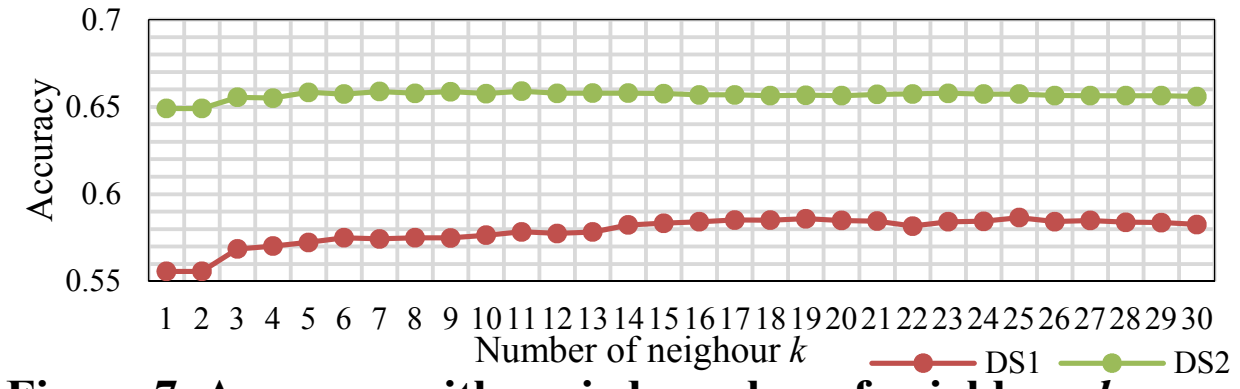

**Figure 7. Accuracy with varied number of neighbors *k*.**

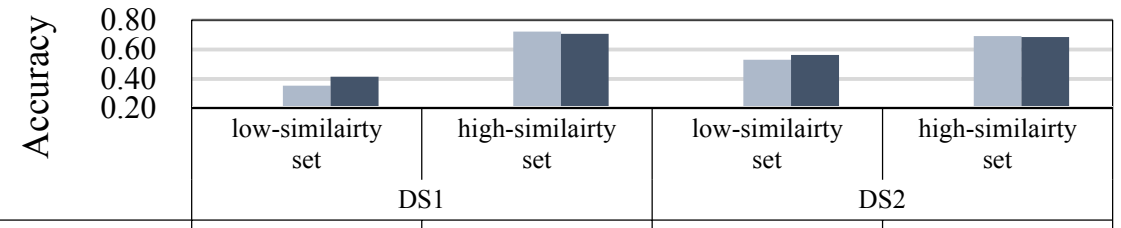

| | DS1 low-similairty set | DS1 high-similairty set | DS2 low-similairty set | DS2 high-similairty set |
|---|---|---|---|---|
| # of test logs | 3309 | 4047 | 1702 | 4855 |
| Accuracy before | 0.354 | 0.721 | 0.530 | 0.691 |
| Accuracy after | 0.416 | 0.707 | 0.563 | 0.684 |

**Figure 8. Accuracy changes before/after KNN strategy.**

#### 1) Influence of Parameters

CAM requires a target value $t$ to determine the cause-specific threshold for each type of cause. Meanwhile, for the new test logs in low-similarity set, CAM employs KNN to analyze their causes. The influences of these parameters are investigated in this subsection.

**Target value *t***. We evaluate the relationship between CAM's accuracy with respect to different target value $t$. To tune the parameter $t$, we set the number of neighbors $k$ to a fixed value ($k$ = 15 in this experiment), and vary $t$ from 0 to 1 with a step size of 0.1. When $t$ = 1, the high-similarity set is empty. In contrast, the low-similarity set is empty when $t$ = 0. Two curves in Fig. 6 show that as $t$ is small (t < 0.4), the accuracy is low. When $t$ increases, the accuracy increases as well. The stable ranges are slightly different: CAM's accuracy is stable when $t$ ranges from 0.6 to 0.9 on DS1, while the accuracy is more stable on DS2 when t varies. We set $t$ = 0.7 in the following experiments.

**Number of neighbors $k$**. To investigate CAM's accuracy with respect to varied $k$, we set $t$ = 0.7 and $k$ varies from 1 to 30 with a step size of 1. When $k$ = 1, it means CAM conducts the prediction only with the top-1 cause. The tendencies of the curves in Fig. 7 are similar to that in Fig. 6. As $k$ is small ($k < 3$), the accuracy is low, and then the accuracy gradually rises to be stable along with the growth of $k$. The accuracy turns to be stable when $k > 14$ on DS1 and $k > 2$ on DS2. We set $k$ = 15 in the following experiments. We do not set an extremely large $k$ to avoid introducing noisy neighbors.

*2) Deep Analysis of Parameter Setting*

We investigate why such parameters ($t$=0.7, $k$=15) work for cause prediction. The target value $t$ splits the new test logs into the high-similarity and low-similarity sets. The value of $k$ controls the prediction strategy. Fig. 8 shows the total size of both high-similarity and low-similarity sets on the two datasets. The light and dark blue bars show the accuracy before and after applying the KNN strategy, respectively.

In Fig. 8, CAM successfully splits the new test logs into two sets. The high-similarity set of DS1 and DS2 both cover more than 50% of the new test logs, in which CAM achieves an accuracy of 72.1% and 69.1% respectively. While the accuracies of the low-similarity sets are only 35.4% and 53.0% for the two datasets. After deciding the low-similarity sets, KNN strategy improves the accuracy of these sets by up to 6.2%, namely, from 35.4% to 41.6% on DS1 and from 53.0% to 56.3% on DS2. However, if we also apply KNN to the high-similarity set, the accuracy drops on both datasets. The reason is that according to the hypothesis of CAM, when the top-1 $Sim_{log}$ is greater than the threshold, the top-1 cause is likely to be the ground-truth cause of the new test alarm.

**Answer to RQ2**. We set $t$ = 0.7 and $k$ = 15 in this study. CAM achieves an accuracy around 70% for the high-similarity set. KNN improves the accuracy in the low-similarity set by up to 6.2%.

### C. Answer to RQ3

*1) Accuracy and AUC Evaluation*

We summarize the experimental results in terms of Accuracy in Fig. 9. CAM achieves an accuracy of 58.3% and 65.8% on the two datasets. It outperforms the baseline algorithms by up to 7.3% on DS1 and 13.3% on DS2.

Out of the three baseline algorithms, no one is superior to the others, since LAC performs well on DS1, while loses its dominance to BFT on DS2. Comparing with LAC, which mines local patterns from test steps, CAM shows its robustness over different datasets as it compares test logs from an overall perspective. Since a random prediction for more than 6 classes can only achieve an accuracy below 16%, the accuracy of CAM shows its ability in cause analysis.

We also present the accuracy with respect to the testing days in Fig. 10 and Fig. 11. Out of the 39 and 21 predicted testing days, CAM performs best on 23 and 17 of them respectively. CAM achieves an accuracy over 80% for more than one-third testing days, namely 14 out of 39 on DS1 and 11 out of 22 on DS2.

We introduce the paired Wilcoxon signed rank test to explore the statistical significance between the performance of CAM and baseline algorithms. The p-values are 0.002, 0.013, 0 on DS1 and 0.003, 0.002, 0.001 on DS2, when comparing the accuracy of CAM against LAC, BFT, and TM, respectively. Therefore, CAM is superior to the baseline algorithms in terms of accuracy with $p < 0.05$.

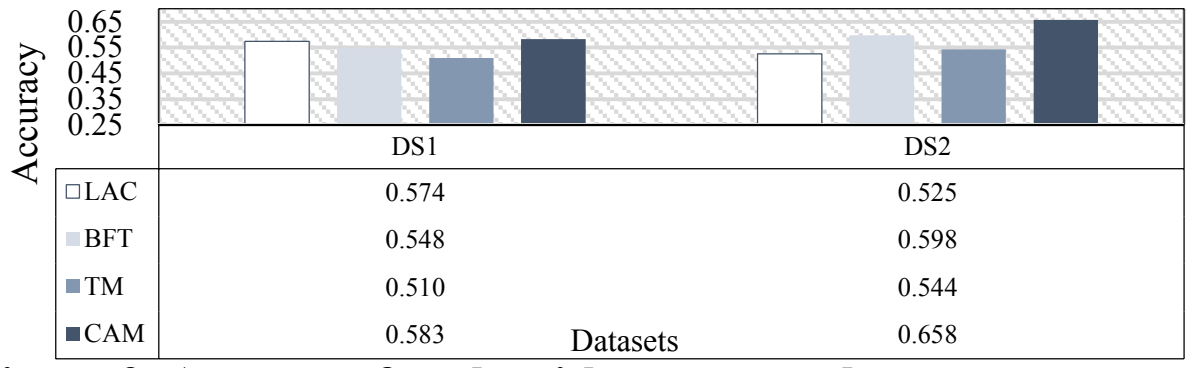

**Figure 9. Accuracy for algorithms on two datasets.**

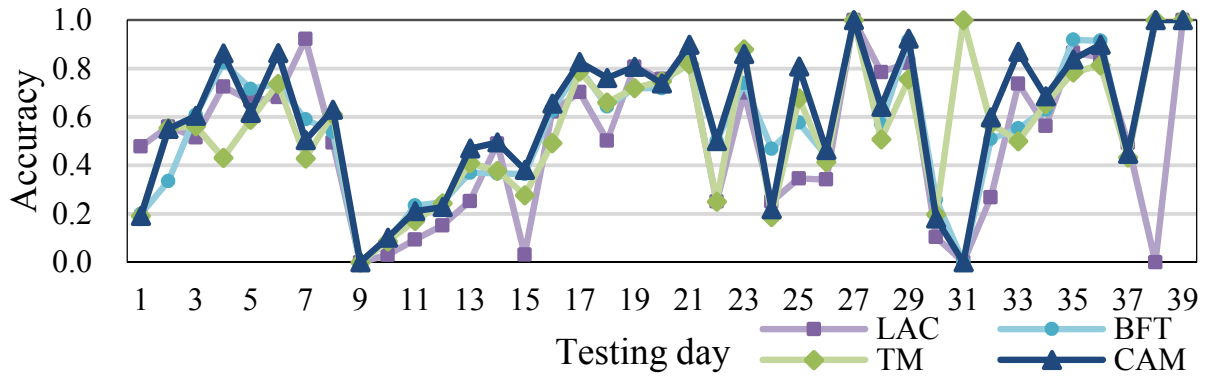

**Figure 10. Accuracy per testing day on DS1.**

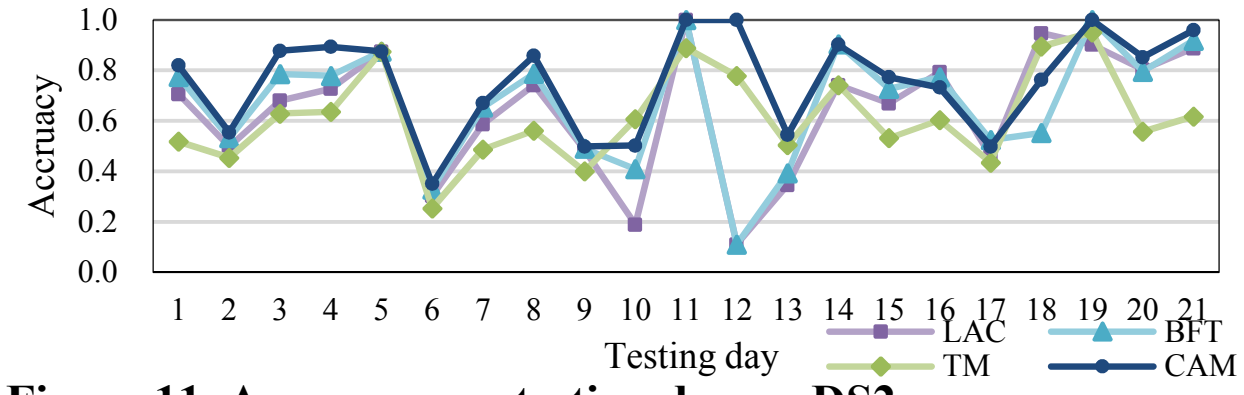

**Figure 11. Accuracy per testing day on DS2.**

**Table 3. Comparison on AUC**

| Algorithm \ Cause | | C1 | C2 | C3 | C4 | C5 | C6 | C7 |
|---|---|---|---|---|---|---|---|---|
| DS1 | LAC | 0.61 | 0.57 | 0.48 | 0.52 | 0.50 | 0.33 | 0.51 |
| | BFT | 0.73 | 0.65 | **0.66** | 0.60 | **0.77** | 0.40 | **0.70** |
| | TM | 0.68 | 0.67 | 0.56 | 0.58 | 0.62 | 0.50 | 0.54 |
| | CAM | **0.77** | **0.71** | 0.59 | **0.61** | 0.62 | 0.50 | 0.62 |
| DS2 | LAC | - | 0.60 | 0.53 | 0.64 | **0.63** | **0.83** | 0.73 |
| | BFT | - | 0.67 | 0.65 | 0.70 | 0.60 | 0.77 | 0.86 |
| | TM | - | 0.62 | 0.51 | 0.68 | 0.52 | 0.77 | 0.78 |
| | CAM | - | **0.68** | **0.66** | **0.81** | 0.51 | 0.74 | **0.87** |

AUC values of the algorithms are presented in Table 3. A value in bold denotes a result, which is better than the other algorithms on the same failure cause. Table 3 shows that some failure causes are difficult to be discovered, e.g., *Environment issue* (C6), since no algorithm is superior to a random classifier on DS1. CAM and BFT have similar performance, which outperforms the other algorithms on 3 out of 7 types of causes on DS1. On DS2, CAM performs better on the majority of causes than the baseline algorithms. For some causes, CAM has an AUC value beyond 0.80, showing that it could provide excellent discrimination on these causes.

*2) Computation Resources Evaluation*

**Table 4. Comparison on computation resources consumption**

| Algorithm | Time (in minutes) | | | | | | Memory | |
|---|---|---|---|---|---|---|---|---|
| | DS1 (7356 test logs) | | | DS2 (6557 test logs) | | | DS1 | DS2 |
| | Training | Test | Total | Training | Test | Total | | |
| LAC | 11.4 | 1 | 12.4 | 3.6 | 1.4 | 5 | 3 GB | 3 GB |
| BFT | 208.6 | 0.3 | 208.9 | 46.8 | 0.2 | 47 | 22 GB | 20 GB |
| TM | 75.1 | 2.8 | 77.9 | 142 | 4.3 | 146.3 | 8 GB | 5 GB |
| CAM | 0 | 6.9 | 6.9 | 0 | 14.4 | 14.4 | 4 GB | 4 GB |

We compare the computation time and memory consumption in this subsection. To compare the computation time, we directly allocate 22GB memory for the algorithms (2GB is reserved for the operating system). Under the incremental framework, the computation time is calculated as the sum of the time for training or testing models after each incremented testing day. To compare the memory consumption, we allocate the memory increasingly from 2GB to 22GB with a step size of 1GB and observe the minimal memory for algorithms to accomplish the prediction.

In Table 4, the columns refer to the algorithm name, the computation time, and the memory consumption on the two datasets. For each dataset, we present the training time, testing time, and total time in the sub-columns of *Time*. CAM conducts testing without spending time on training models. As shown in Table 4, LAC runs extremely fast as it only mines local patterns of test steps. BFT takes the longest time on DS1, since it frequently conducts garbage collection even with 22GB memory. When comparing CAM against BFT and TM in terms of total time, CAM runs nearly 3 to 30 times faster than BFT, and 10 times faster than TM. CAM takes 6.9 and 14.4 minutes in analyzing 7356 and 6557 test logs over the two datasets respectively, which means that it takes only 0.06s to 0.13s on average in helping testers analyze one test log.

The training time is the main overhead for most algorithms. For example, it takes 208.6 minutes for BFT to incrementally train models on 7356 test logs. As a result, when the test logs continue increasing, it may take days to update models. In contrast, a new test log can be immediately updated to CAM, once the tester verifies the cause of that log.

For memory consumption, most algorithms consume no more than 8GB for prediction, except BFT. BFT cannot complete predictions until we allocate 22GB memory. In contrast, CAM only takes less than 4GB memory. The underlying reason is that CAM conducts prediction without the need of allocating huge memory to train models.

In addition, along with the growth of historical test logs, we can set a time frame of historical test logs to limit the computation time and memory consumption of CAM.

**Answer to RQ3**. CAM outperforms the baseline algorithms over distinct evaluation metrics. Meanwhile, CAM is resources saving as it takes about 0.1s and less than 4GB memory to process a test log.

### *D. Answer to RQ4*

**Table 5. AUC values for CAM and CAM-FP**

| Algorithm \ Cause | | C1 | C2 | C3 | C4 | C5 | C6 | C7 |
|---|---|---|---|---|---|---|---|---|
| DS1 | CAM-FP | 0.73 | 0.70 | 0.59 | 0.57 | 0.59 | 0.50 | 0.62 |
| | CAM | **0.77** | **0.71** | 0.59 | **0.61** | **0.62** | 0.50 | 0.62 |
| DS2 | CAM-FP | - | 0.67 | **0.76** | 0.76 | **0.52** | 0.67 | 0.84 |
| | CAM | - | **0.68** | 0.66 | **0.81** | 0.51 | **0.74** | **0.87** |

**Table 6. Accuracy, total time, and memory for CAM and CAM-FP**

| Algorithm | DS1 | | | DS2 | | |
|---|---|---|---|---|---|---|
| | Accuracy | Total time | Memory | Accuracy | Total time | Memory |
| CAM-FP | 0.555 | 39.2 min | 4GB | 0.634 | 46.4 min | 4GB |
| CAM | 0.583 | 6.9 min | 4GB | 0.658 | 14.4 min | 4GB |

CAM utilizes function points to conduct historical test log selection. In a well-planned testing process, the function points are predefined when testers develop test scripts [45]. However, function points may be unavailable, if testers do not organize test scripts with them in some projects. We propose an algorithm named CAM-FP to simulate such a situation. CAM-FP searches among all the historical test logs to conduct prediction without historical test log selection.

We show the AUC, accuracy, total time, and memory for CAM-FP and CAM over the two datasets in Table 5 and Table 6. We find that historical test log selection could remove considerable noisy test logs, since CAM outperforms CAM-FP by 2.4%-2.8% in terms of accuracy. CAM also outperforms CAM-FP on the majority of causes in terms of AUC. Besides, historical test log selection can significantly shorten the running time of CAM. After selection, CAM shortens the running time from 39.2 to 6.9 on DS1 and from 46.4 to 14.4 on DS2 in minutes.

We find that although the function points of test scripts are removed, CAM-FP still achieves competitive performance against the baseline algorithms. For several types of causes, the AUC value of CAM-FP is equal to or slightly better than CAM. It shows the robustness of CAM on different situations.

**Answer to RQ4**. Historical test log selection reduces the noisy test logs and shortens the running time for CAM. Without the function points of test scripts, CAM still achieves competitive performance.

### *E. Answer to RQ5*

We integrate CAM into the automatic testing system of our industry partner. CAM achieves an average accuracy of 72% after two months of running. "*This version (CAM) is better than the intelligent analysis tool of last version (manually building regular expressions).*" An interesting finding is that CAM performs better in a real development scenario than in the experiments. The reason is that testers tend to conduct software testing several rounds per day. The causes between each round may be similar. After testers verify the causes in a round, CAM immediately utilizes the information of the test logs to predict the causes in the next round. However, it is hard to decide the round for each test log in the experiment. When we split the test logs according to the testing days, it may limit the information available to CAM.

In addition, instead of simply presenting the causes for test alarms, CAM also presents the differences between test logs. "*I think CAM is accurate. Actually, I will not believe in an automatic tool. However, after presenting the historical test logs, I can quickly decide whether the prediction is correct. CAM accelerates my work.*" The presentation of results is important since good presentations may make the prediction easy to comprehend [27]. Such human factor, namely, the influence of different presentations, is not the focus of this study. We leave it as a future work.

Some testers also suggest new features for CAM, including "*labeling the defect-related snippets from the lengthy test logs*", "*provide suggestions on how to fix different types of defects*", etc. These features drive us to continue improving CAM.

**Answer to RQ5**. CAM achieves an average accuracy of 72% after two months of running in a real development scenario. CAM accelerates testers' work with accurate result and comprehensible presentation.

## VI. THREATS TO VALIDITY

### A. Experiment Construction

The generality of the cause prediction algorithm in CAM should be further studied, since the algorithms may be sensitive to the datasets. To alleviate this threat, we evaluate CAM over two industry datasets with more than 14,000 test logs and deploy it in a real development scenario.

As the ground-truth causes of the test alarms are manually labeled by testers, there may induce some errors. However, industries have a strict process to control the quality of software testing. The error rate is usually under control.

### B. Method Construction

In this study, the quality of test logs may influence the results of CAM. Currently, researchers study how to automatically decide where to log and what to log for developers [29]. These techniques may improve the quality of test logs for CAM to conduct the prediction.

In addition, we mainly focus on analyzing test alarms with test logs. In order to detect the exact causes of test alarms, testers may go through various software artifacts, e.g., test logs, test script codes, etc. The accuracy of CAM may be further improved by leveraging more software artifacts. However, test logs are still an effective debugging tool [20].

## VII. RELATED WORK

### A. Test Alarm Classification

Rogstad et al. [48] distinguish test code obsoleteness from regression faults for database applications with a set of attributes, e.g., table names, SQL statements, etc. Hao et al. [34] classify test alarms into product code defect and obsolete test at the unit testing stage with complex attributes related to test complexity and program execution measurements. However, these techniques are either unique to database products or expensive to collect complexity information [4] in large software systems.

Herzig and Nagappan classify test alarms in SIT [4]. They detect all false test alarms in Microsoft products with association rules, since the number of false test alarms is a measurement to measure test quality [13]. Different from detecting false test alarms, CAM analyzes the test alarms causes, which is more complex than the binary classification.

Recently, several patents are filed to construct systems to analyze test results [14], bucket failure messages [5], and analyze error logs with regular expression [9]. However, technical details and evaluations are not presented in these patents. As to a survey with testers in our industry partner, they manually build regular expressions to classify test alarms and achieve an accuracy of 20%-30% over distinct projects.

### B. Log Analysis

Previous work mainly analyzes two types of logs, namely, the logs generated by the released software product (system log) and the logs generated in the testing activity (test log).

For the system log analysis, Oliner et al. discuss the advances and challenges in system log analysis [20]. Shang et al. [19] conduct program verification for big data analytic applications with test logs. Fu et al. [15] and Xu et al. [16] conduct anomaly detection through log analysis. Besides, system logs are also used to diagnose the underlying causes of system anomaly [17] [18]. However, such logs may lack information to analyze the defects in testing activities.

For the test log analysis, previous work mainly utilize test logs to solve the oracle problem. Oracle problem is to check whether a test result reveals a failure or not. Andrews et al. [10] and Tu et al. [11] analyze test logs for oracle problem with state-machine-based models. Yantzi et al. [12] conduct an industrial evaluation of methodologies for oracle problem. Recent work by Anderson et al. [49] constructs attributes from test logs to predict the oracle.

In conclusion, system logs analysis is a post-process of software testing. Our work falls into test log analysis. Instead of solving the oracle problem, we predict the underlying causes for test alarms.

### C. Failure Clustering

Most studies in failure clustering detect failures in product code with execution profiles. Execution profiles capture the execution of basic blocks and conditional branches of software. Liu et al. [21] cluster failures with fault location techniques. Dickson et al. [22] cluster program executions and identify failures among the clusters with unusual-profile hypothesis. DiGiuseppe et al. [24] utilize latent-semantic-analysis techniques for more precise failure clustering.

Besides, Podgurski et al. [23] train pattern classifiers to group similar failures in product code. Francis et al. [25] refine the failure cluster results with two tree-based techniques. Lo et al. [26] capture program execution profiles to train machine learning models for identifying failures in software product.

Clustering failures in product codes is a subsequent process of test alarm analysis. Approaches in analyzing product codes may fail to identify the failure causes in software testing. Meanwhile, execution profiles are hard to collect for testers in SIT. Different from clustering failures in product code, we classify the causes of test alarms in SIT.

## VIII. CONCLUSION AND FUTURE WORK

In this study, we present our attempt towards predicting the multiple causes for test alarms in SIT. Our model leverages the test logs of historical test alarm to analyze the new test alarm. We evaluate our model over two industrial projects with more than 14,000 test alarms. Our model shows an accuracy of 58.3% and 65.8%, respectively. We deploy CAM for our industry partner and achieve an accuracy of 72% after two months of running, nearly three times better than their previous strategy with regular expressions. Our technique provides a direction for industry to analyze test alarms. In the future, we plan to employ more software artifacts to improve CAM and verify CAM over more software projects.

## ACKNOWLEDGMENT

We greatly thank our industrial companion for sharing their datasets for research. This work is supported in part by the New Century Excellent Talents in University under Grant NCET-13-0073, in part by the Fundamental Research Funds for the Central Universities under Grant DUT14YQ203, and in part by the National Natural Science Foundation of China under Grants 61370144, 61403057 and 61502345.